\documentstyle[epsfig,longtable]{aipproc}

\begin{document}
\title{Measurement of $\Lambda^0$  polarization in $\nu_\mu$ CC 
interactions in NOMAD}
\author{Dmitry V. Naumov\\
for the NOMAD Collaboration}
\address{Laboratory of Nuclear Problems, Joint Institute for Nuclear Research, 141980, Dubna, Russia\\
e-mail: \it naumov@thsun1.jinr.ru}
\maketitle
\begin{abstract}
The $\Lambda^0$ polarization in $\nu_\mu$ charged current interactions
has been measured in the NOMAD experiment.
The event sample (8087 reconstructed $\Lambda^0$'s)
is more than an order of magnitude larger
than that of previous bubble chamber experiments, while the
quality of event reconstruction is comparable.
We observe negative polarization along the $W$-boson direction which
is enhanced in the target fragmentation region:
$P_x (x_F < 0) = -0.21 \pm 0.04 \mbox{(stat)} \pm 0.02 \mbox{(sys)}$.
In the current fragmentation region we find
$P_x (x_F > 0) = -0.09 \pm 0.06 \mbox{(stat)} \pm 0.03 \mbox{(sys)}$.
These results provide a test of different models describing the
nucleon spin composition and the spin transfer mechanisms.
A significant transverse polarization
(in the direction orthogonal to the $\Lambda^0$ production plane)
has been observed for the first time in a neutrino experiment:
$P_y = -0.22 \pm 0.03 \mbox{(stat)} \pm 0.01 \mbox{(sys)}$.
The dependence of the absolute value of $P_y$ 
on the $\Lambda^0$ transverse momentum
with respect to the hadronic jet direction is in qualitative agreement
with the results from unpolarized hadron-hadron experiments.
\end{abstract}

\vspace*{-0.7cm}
\section{Introduction}
A study of the $\Lambda^0$ polarization in (anti)neutrino nucleon DIS 
is motivated by several reasons. First, a possible longitudinal
polarization in the {\it target fragmentation region} can be related to
the polarized nucleon strangeness which is a conclusion (in the 
framework of quark parton model assuming $SU(3)_F$ symmetry and vanishing
gluon spin contribution to the nucleon spin) of DIS experiments with both
polarized beam and target ~\cite{EMC}. The authors ~\cite{PSM} attempt
to explain the negative sign of the nucleon strangeness and predict that 
it could manifest itself as a {\it negative} (aligned in the opposite direction
to the $W$ exchange boson) longitudinal polarization of $\Lambda^0$ hyperons
produced in (anti)neutrino nucleon DIS. Second, a measurement of the 
longitudinal polarization  in the {\it current fragementation region} 
provides a test of different models for the $\Lambda^0$ spin 
structure ~\cite{cfr,cfr_aram,cfr_soffer} with a {\it clean flavour separation} provided by 
the nature of neutrino interactions. Last but not least,  transverse
polarization of $\Lambda^0$ hyperons has been observed for a long time
in unpolarized hadron-hadron experiments  ~\cite{lambda_hadron},
and was never  observed in (anti)neutrino nucleon DIS experiments 
~\cite{lambda_neutrino}. This surprizing feature challenges experimental and
theoretical efforts in this field. The neutrino nucleon DIS is again 
exceptional due to the different $\Lambda^0$ production mechanisms in the 
target and current fragmentation regions in contrast to $pp$ scattering.
The quark and di-quark fragmentations are believed to be the subject of 
the current and target  fragmentation regions respectively. Therefore, it is
possible to study the transverse polarization in connection with the different
mechanisms of  $\Lambda^0$ production.
\vspace*{-0.5cm}
\section{Experimental procedure and analysis}
The active part of the NOMAD detector consists of 44 drift chambers located
in a $0.4$ Tesla magnetic field. The drift chambers serve as a nearly 
isoscalar target for neutrino interactions and as a tracking medium.
These drift chambers provide an overall efficiency for charged track
reconstruction of better than 95\% and a momentum resolution of 
approximately 3.5\% in the momentum range of interest (less than 
10~$\mbox{GeV}/\mbox{c}$). Reconstructed tracks are used to determine 
the event topology (the assignment of tracks to vertices), to reconstruct 
the vertex position and the track parameters at each vertex and, finally, to
identify the vertex type (primary, secondary, V$^0$, etc.).

$\Lambda^0$ hyperons appear in the detector as two charged tracks
with opposite charges emerging from a common vertex separated from 
the primary interaction vertex (V$^0$ -like signature). These events 
correspond to $\Lambda^0 \to p \pi^-$ decay. The background to $\Lambda^0$ 
decays consists of K$^0_S \to \pi^+ \pi^-$, 
${\bar \Lambda}^0 \to \bar p \pi^+$ decays, $\gamma \to e^+ e^-$ conversions,
and random combinations of tracks  wrongly labeled as  V$^0$s. To identify 
$\Lambda^0$ hyperons we  first  apply some quality cuts to reject as much 
as possible of the combinatorial background, $\gamma$-conversions, and 
secondary interactions, we then perform a kinematic fit
with energy and momentum constraints for each V$^0$ for the final resolution
of V$^0$-like particles. As a result we obtained 8087 reconstructed and
identified $\Lambda^0$ hyperons with about $4\%$ background 
contamination in our data sample ~\cite{NOMAD_lambda}.
This sample is used for the polarization analysis reported below.

The $\Lambda^0$ polarization is measured through the {\em asymmetry} 
in the angular distribution of the protons in the parity violating
decay process $\Lambda^0 \to p \pi^-$.
In the $\Lambda^0$ rest frame the decay protons are distributed as:
$
\frac{1}{N}\frac{d N}{d \Omega} = 
\frac{1}{4\pi}(1+\alpha_\Lambda \mathbf P \cdot \mathbf k),
\label{eq:asymmetry}
$
where $\mathbf P$ is the $\Lambda^0$ polarization vector,
$\alpha_\Lambda = 0.642 \pm 0.013$~\cite{PDG} is the decay asymmetry parameter 
and $\mathbf k$ is the unit vector along the decay proton direction.
A fit of the raw angular distributions of the decay protons in the data can 
only be performed after correction for the detector acceptance. To take
into account the detector acceptance, and smearing of the angular distributions
we developed a new 3-dimensional method for the polarization analysis
~\cite{NOMAD_lambda}. We used its 1-dimensional option for the results 
reported below because of its better applicability to the samples with 
low statistics as is the case in the study of the polarization dependence
on different kinematic variables. 

The axes are defined as follows (in the $\Lambda^0$  rest frame):
\begin{itemize}
\item $\mathbf n_x =  e_W$, where $\mathbf e_W$ is the reconstructed 
$W$-boson direction;
\item $\mathbf n_y = \mathbf e_W \times \mathbf e_{T} /
|\mathbf e_W \times \mathbf e_{T} |$ axis is orthogonal to the $\Lambda^0$ 
production plane 
\item $\mathbf n_z = \mathbf n_x \times \mathbf n_y$. 
\end{itemize}
\vspace*{-1cm}

\section{Results and discussion}
\vspace*{-0.2cm}

Table ~\ref{tab:general} displays the results for the polarization of $\Lambda^0$ 
hyperons in our sample as a function of $x_F$. We observe
negative longitudinal (``$P_x$'') and transverse (``$P_y$'') polarizations 
of $\Lambda^0$'s which are enhanced
in the target fragmentation region. Note that transverse polarization has never 
been observed before in (anti)neutrino nucleon DIS experiments. It is believed that 
the origin of $\Lambda^0$ polarization is different in the target and in the current
fragmentation regions.
Therefore it can be useful to study  $\Lambda^0$ polarization at $x_F<0$ and $x_F>0$
separately.
\vspace*{-0.5cm}

\begin{table}[htb]
\begin{center}
\caption{\label{tab:general} \it Dependence of the $\Lambda^0$ polarization on $x_F$ in $\nu_\mu$ CC events 
(statistical errors only).}
\begin{tabular}{||c|c|c|c|c|c||}
\hline
\hline
 & & & \multicolumn{3}{|c||}{$\Lambda^0$ Polarization} \\
\cline{4-6}
Selection & Entries & $<x_F>$ & $P_x$         &  $P_y$         &$P_z$ \\
\hline
\hline
full sample & 8087 & $-0.18$ & $-0.15 \pm 0.03$ & $-0.22 \pm 0.03$ & $-0.04 \pm 0.03$\\
\hline
\hline
$x_F<0$ & 5608 & $-0.36$ & $-0.21 \pm 0.04$ & $-0.26 \pm 0.04$ & $-0.08 \pm 0.04$\\
$x_F>0$ & 2479 & $0.21$ & $-0.09 \pm 0.06$ & $-0.10 \pm 0.06$ & $ 0.02 \pm 0.06$\\
\hline 
\hline
\end{tabular}
\end{center}
\end{table}

\vspace*{-1.2cm}

\subsection{Target fragmentation region}
\vspace*{-0.7cm}

\begin{figure}[htb]
\begin{minipage}[r]{0.48\textwidth}
  \centering\epsfig{file=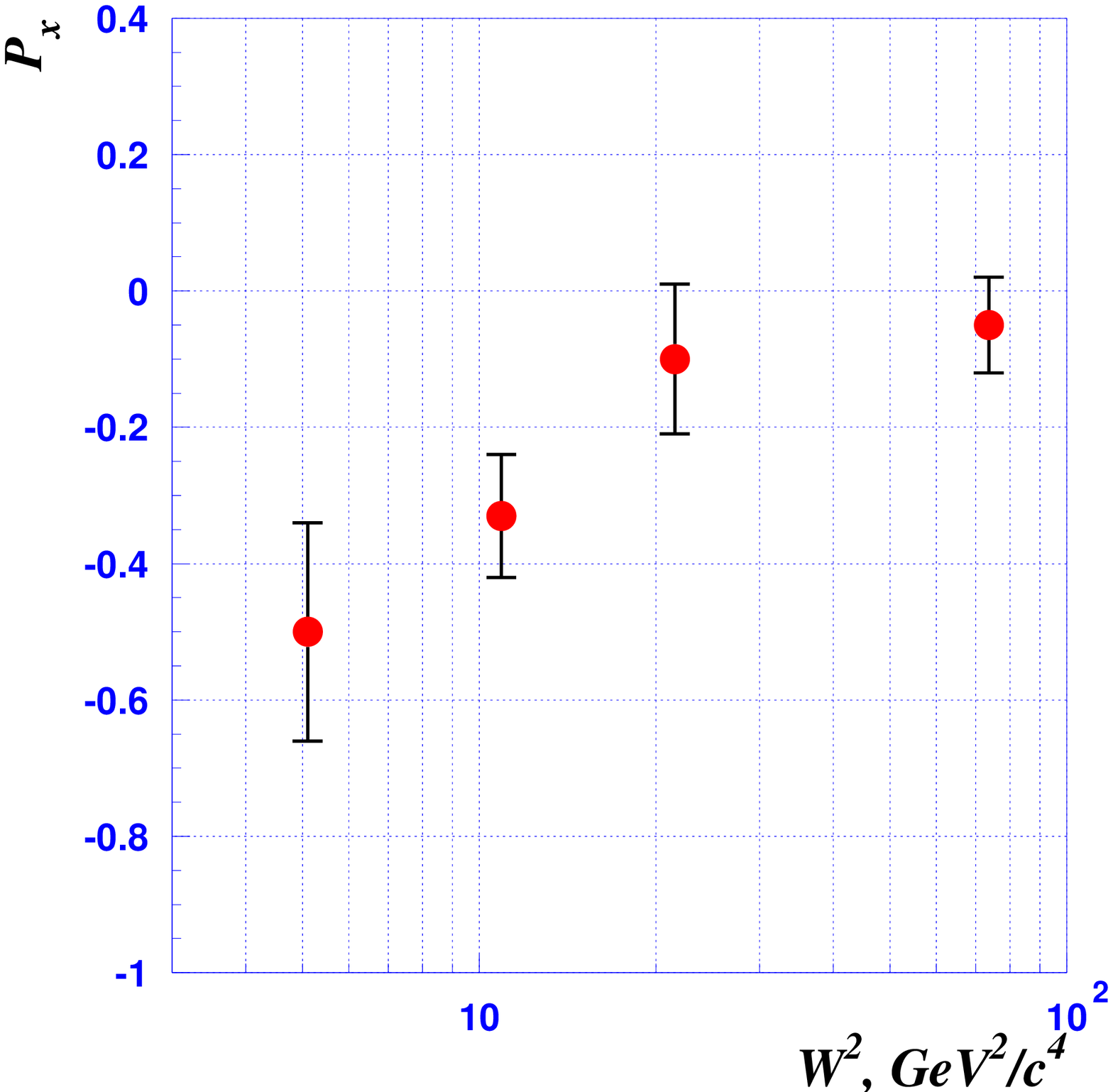,height=4.7cm,width=\textwidth} 
  \caption{\label{fig:long_w2_xfn} Longitudinal polarization of $\Lambda^0$ as a function of $W^2$ for $x_F<0$}
\end{minipage}%
\hspace*{0.02\textwidth}
\begin{minipage}[l]{0.48\textwidth}
  \centering\epsfig{file=EPS/cosy-pt2-xfn.epsi,height=4.7cm,width=\textwidth} 
  \caption{\label{fig:trans_pt_xfn} Transverse polarization of $\Lambda^0$ as a function of $P_T$ for $x_F<0$}
  \end{minipage}
\end{figure}

\vspace*{-1.cm}

~\subsubsection{Longitudinal polarization}

The dependence of the longitudinal polarization of $\Lambda^0$ on $W^2$ at $x_F<0$
is shown in Fig.~\ref{fig:long_w2_xfn}. Large negative $P_x$ is observed at small $W^2$, while at 
larger $W^2$ the longitudinal polarization vanishes. 
Such an effect can be interpreted as a manifestation of 
the polarized nucleon strangeness due to the larger probability for $\Lambda^0$ at small $W^2$ to
include an $s-$quark originally present in the nucleon, while at larger $W^2$ the $s-$quarks (presumably
unpolarized) are also created in the fragmentation process. The same dependence of $P_x$ on $Q^2$
is observed.

\newpage
\subsubsection{Transverse polarization}

We have performed a study of the dependence of the transverse polarization on  
the $\Lambda^0$ transverse momentum with respect to the jet direction  ($p_T$) in the target
fragmentation region and found it to be in qualitative agreement (both sign and shape) with that found in 
unpolarized hadron-hadron collisions ~\cite{lambda_hadron}. Also, we observed no dependence
of $P_y$ on $W^2$. These features make possible to conclude that the origin of the transverse polarization
is in the fragmentation process.

\vspace*{-0.5cm}
\subsection{Current fragmentation region}
\begin{figure}[htb]
\begin{minipage}[r]{0.6\textwidth}
\centering\epsfig{file=EPS/kotz_soff_moriond.epsi,height=5cm,width=\textwidth} 
\caption{\label{fig:long_z_xfp} Longitudinal polarization of $\Lambda^0$ as a function of $z$ in comparison
with predictions from [4] (left) and [5] (right)}
\end{minipage}%
\hspace*{0.02\textwidth}
\begin{minipage}[l]{0.36\textwidth}
\centering\epsfig{file=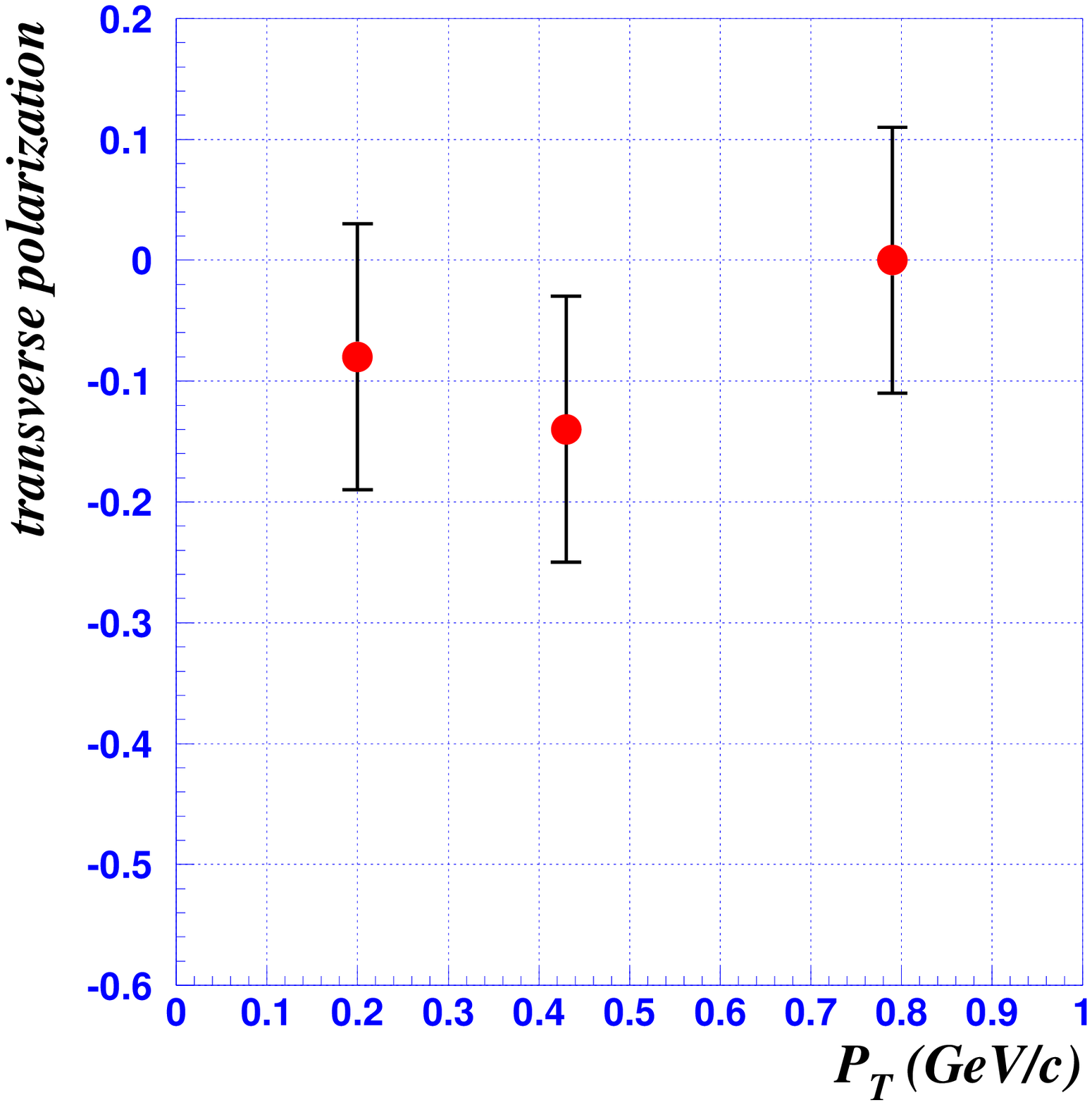,height=5cm,width=\textwidth} 
\caption{\label{fig:trans_pt_xfp} Transverse polarization of $\Lambda^0$ for $x_F>0$ as a function of $P_T$}
\end{minipage}
\end{figure}
\vspace*{-1cm}
~\subsubsection{Longitudinal polarization}
Measurement of the longitudinal $\Lambda^0$ polarization in the current fragmentation region provides
a test of different models for the $\Lambda^0$ spin structure. A comparison of our data to theoretical
calculations ~\cite{cfr_aram} and ~\cite{cfr_soffer} performed for different models of the $\Lambda^0$
spin content (details can be found in  ~\cite{cfr_aram,cfr_soffer}) 
is presented in Fig.~\ref{fig:long_z_xfp}. One can draw the conclusion that naive quark parton model 
is favoured by our measurement while still some progress in theoretical calculations is expected.
\vspace*{-0.5cm}
\subsubsection{Transverse polarization}
Transverse polarization of $\Lambda^0$ in the current fragmentation region in neutrino-nucleon DIS
is related to the quark fragmentation processes, therefore it is crucial to look for its $p_T$ dependence.
Fig.~\ref{fig:trans_pt_xfp} displays such a dependence. 
\vspace*{-0.5cm}
\subsection{Target nucleon effects}
Imposing a cut on the total charge ($Q_{tot}$) of the event we can study $\Lambda^0$ polarization on different 
target nucleons. We select $\nu_\mu p$ ($\nu_\mu n$)-like events requiring $Q_{tot} \ge 1$  ($Q_{tot} \le 0$)
with purity of the selection 76\% (85\%).
The results are summarized in Table.~\ref{tab:target}. There is a strong dependence
of the polarization vector on the  target nucleon. We attribute it to the different contribution of
the polarization transfer from $\Sigma^*,\Xi,\Sigma^0$ to $\Lambda^0$ during their decays into $\Lambda^0$
in the final state.
\vspace*{-0.5cm}
\begin{table}[htb]
\begin{center}
\caption{\label{tab:target}\it The dependence of the $\Lambda^0$ polarization on the type of target nucleon.}
\begin{tabular}{||c|c|c|c|c||}
\hline
\hline
 & & \multicolumn{3}{|c||}{$\Lambda^0$ Polarization} \\
\cline{3-5}
Target & Entries & $P_x$         &  $P_y$         &$P_z$ \\
\hline
\hline
``proton'' & 3472 & $-0.26 \pm 0.05$ & $-0.09 \pm 0.05$ & $-0.07 \pm 0.05$\\
$x_F<0$    & 2407 & $-0.29 \pm 0.06$ & $-0.10 \pm 0.06$ & $-0.09 \pm 0.06$\\
$x_F>0$    & 1065 & $-0.23 \pm 0.09$ & $-0.06 \pm 0.09$ & $-0.02 \pm 0.10$\\
\hline 
\hline
``neutron''& 4615 & $-0.09 \pm 0.04$ & $-0.30 \pm 0.04$ & $-0.03 \pm 0.05$\\
$x_F<0$    & 3201 & $-0.16 \pm 0.05$ & $-0.37 \pm 0.05$ & $-0.07 \pm 0.05$\\
$x_F>0$    & 1414 & $ 0.01 \pm 0.08$ & $-0.11 \pm 0.08$ & $ 0.04 \pm 0.09$\\
\hline 
\hline
\end{tabular}
\end{center}
\end{table}
\vspace*{-1cm}
\section{conclusion}
\vspace*{-0.2cm}
Our measurements indicate for many interesting phenomena are hidden in the nucleon and in the 
fragmentation process. 
\vspace*{-0.5cm}
\section*{Acknowledgements}
\vspace*{-0.2cm}
Many thanks to the SPIN2000 organizers for making possible me to participate in the symposium.
Special thanks to the NOMAD collaborators making possible this work to appear.

\end{document}